\documentclass[aps,prl,twocolumn,showpacs,groupedaddress]{revtex4} 
\usepackage{graphicx} 
\usepackage{dcolumn}   
\usepackage{bm}       
\usepackage{amssymb}  
\usepackage{bm}
\begin{document}

\hspace{5.2in} \mbox{Fermilab-Pub-08/335-E}
\title{Observation of the doubly strange \boldmath $b$ \unboldmath baryon \boldmath $\Omega_b^-$ \unboldmath}
% LIST_OF_AUTHORS_R2.TEX                 7/25/08            
%
\author{V.M.~Abazov$^{36}$}
\author{B.~Abbott$^{75}$}
\author{M.~Abolins$^{65}$}
\author{B.S.~Acharya$^{29}$}
\author{M.~Adams$^{51}$}
\author{T.~Adams$^{49}$}
\author{E.~Aguilo$^{6}$}
\author{M.~Ahsan$^{59}$}
\author{G.D.~Alexeev$^{36}$}
\author{G.~Alkhazov$^{40}$}
% ...

\affiliation{\vspace{0.1 in}(The D\O\ Collaboration)\vspace{0.1 in}}
% ...
\affiliation{$^{6}$University of Alberta, Edmonton, Alberta, Canada,
                Simon Fraser University, Burnaby, British Columbia, Canada,
                York University, Toronto, Ontario, Canada, and
                McGill University, Montreal, Quebec, Canada}
% ...
\affiliation{$^{29}$Tata Institute of Fundamental Research, Mumbai, India}
% ...
\affiliation{$^{36}$Joint Institute for Nuclear Research, Dubna, Russia}
% ...
\affiliation{$^{40}$Petersburg Nuclear Physics Institute,
                St. Petersburg, Russia}
% ...
\affiliation{$^{49}$Florida State University, Tallahassee, Florida 32306, USA}
% ...
\affiliation{$^{51}$University of Illinois at Chicago,
                Chicago, Illinois 60607, USA}
% ...
\affiliation{$^{59}$Kansas State University, Manhattan, Kansas 66506, USA}
% ...
\affiliation{$^{65}$Michigan State University,
                East Lansing, Michigan 48824, USA}
% ...
\affiliation{$^{75}$University of Oklahoma, Norman, Oklahoma 73019, USA}
% ...
 
\date{August 29, 2008}

\begin{abstract}
We report the observation of the doubly strange $b$ baryon $\Omega_{b}^{-}$ in the decay channel $\Omega_b^- \to J/\psi\thinspace\Omega^-$, with $J/\psi\to\mu^+\mu^-$ and $\Omega^-\to\Lambda K^-\to (p\pi^-) K^-$, in $p\bar{p}$ collisions at $\sqrt{s}=1.96$ TeV. Using approximately 1.3 fb$^{-1}$ of data collected with the D0 detector at the Fermilab Tevatron Collider, we observe $17.8\pm 4.9\thinspace({\rm stat.})\pm 0.8\thinspace({\rm syst.})$ $\Omega_b^-$ signal events at a mass of $6.165\pm 0.010\thinspace ({\rm stat.})\pm 0.013\thinspace ({\rm syst.})$ GeV. The significance of the observed signal is $5.4\sigma$, corresponding to a probability of $6.7\times 10^{-8}$ of it arising from a background fluctuation.
\end{abstract}

\maketitle 

The $\Omega^-$ baryon, composed of three strange quarks, played an important historical role in our understanding of the basic structure of matter. Its discovery in 1964~\cite{omega} at a mass predicted from SU(3) symmetry breaking was a great success for the theory~\cite{alltheory}. The $\Omega_b^-\ (bss)$ (charge conjugate states are assumed throughout this Letter) is a predicted heavy cousin of the $\Omega^-$ with a $b$ quark replacing one of the three strange quarks. While the $\Omega^-$ has $J^P=3/2^+$, the ground state $\Omega_b^-$ is expected to have $J^P=1/2^+$,  a mass between $5.94 - 6.12$~GeV and a lifetime such that $0.55<\tau(\Omega_{b}^{-})/\tau(B^{0})<1.10$~\cite{theory}. 

In this Letter, we report the first observation of the $\Omega_b^-$ baryon, fully reconstructed from its decay $\Omega_b^-\to J/\psi\thinspace\Omega^-$, with $J/\psi\to\mu^+\mu^-$, $\Omega^-\to\Lambda K^-$ and $\Lambda\to p\pi^-$. The analysis is based on a data sample of 1.3 fb$^{-1}$ collected in $p\bar{p}$ collisions at $\sqrt{s}=1.96$ TeV with the D0 detector~\cite{d0det} at the Fermilab Tevatron Collider. The detector components most relevant to this analysis are the central tracking system and the muon spectrometer. The central tracking system consists of a silicon microstrip tracker (SMT) and a central fiber tracker (CFT) inside a 2 Tesla superconducting solenoid. The SMT is optimized for tracking and vertexing over the pseudorapidity region $|\eta|<3$ while the CFT has coverage for $|\eta|<2$. A liquid argon and uranium calorimeter provides coverage up to $|\eta|<4.2$. The muon spectrometer covers $|\eta|<2$. 

The $\Omega_b^-\to J/\psi\thinspace\Omega^-\to J/\psi\thinspace\Lambda K^-\to J/\psi\thinspace p\pi^-\thinspace K^-$ decay topology is similar to that of the $\Xi_b^-\to J/\psi\thinspace\Xi^-\to J/\psi\thinspace\Lambda\thinspace\pi^-\to J/\psi\thinspace p\pi^-\thinspace\pi^-$ decay with $\Omega^-$ in place of $\Xi^-$ and $K^-$ in place of $\pi^-$. Consequently, the reconstruction of the $J/\psi$ and $\Lambda$ and their selection discussed below follow closely the analysis that led to the first direct observation of the $\Xi_b^-$ baryon~\cite{d0xib}. However, in this analysis we use a multivariate selection for the $\Omega^-$ owing to the smaller signal to background ratio compared to that for the $\Xi^-$ in the $\Xi_b^-$ analysis. We use the {\sc pythia} Monte Carlo (MC) program~\cite{pythia} to generate $\Omega_b^-$ and the {\sc evtgen} program~\cite{evtgen} to simulate $\Omega_b^-$ decays. The $\Omega_b^-$ mass and lifetime are set to be 6.052~GeV and 1.54~ps respectively. The generated events are subjected to a {\sc geant}~\cite{geant} based D0 detector simulation, and to the same reconstruction and selection programs as the data. We optimize the $\Omega^{-}$ selection using MC $\Omega_b^-$ events for the signal and a sample of $J/\psi(\Lambda K^{+})$ data (referred to below as wrong-sign events) for the background, while keeping the $J/\psi\thinspace\Omega^{-}$ data blinded. Once all selection criteria have been determined, we apply them to the $J/\psi\thinspace\Omega^{-}$ data.

We begin the event selection by reconstructing $J/\psi\to\mu^+\mu^-$ candidates from two oppositely charged muons with transverse momentum ($p_{T}$) greater than 1.5~GeV that are compatible with being from a common vertex. Muons are identified by tracks reconstructed in the central tracking system that are matched with either track segments in the muon spectrometer or calorimeter energy deposits consistent with a minimum ionizing particle. Events must have a well-reconstructed $p\bar{p}$ interaction point that we take to be the $\Omega_{b}^{-}$ production vertex and a $J/\psi\to\mu^+\mu^-$ candidate in the mass window $2.75<M_{\mu\mu}<3.40$~GeV. Events with $J/\psi$ candidates are reprocessed with a version of the track reconstruction algorithm that increases the efficiency for tracks with low $p_T$ and high impact parameters.

We form $\Lambda \to p \pi^-$ candidates from two oppositely charged particles, each with $p_{T} > 0.2$~GeV, that are consistent with having originated from a common vertex. The two tracks are required to have a total of no more than two hits in the tracking system before the reconstructed $p\pi^-$ vertex. The impact parameter significance (the impact parameter with respect to the $p\bar{p}$ interaction point divided by its uncertainty) must exceed four for at least one of the tracks and three for the other. The track with the higher $p_T$ is assumed to be the proton. MC studies show that this assignment leads to the correct combination nearly 100\% of the time. 
Furthermore, we require the $\Lambda$ transverse decay length to be greater than four times its uncertainty and the proper decay length to exceed ten times its uncertainty, where the transverse decay length is the distance between the production and decay vertices in the transverse plane while the proper decay length is the transverse decay length corrected by the Lorentz boost calculated from $p_T(\Lambda)$. $\Lambda$ candidates must have reconstructed masses between 1.108 and 1.126~GeV.

\begin{figure}[htb]
\begin{center}
\begin{tabular}{cc}
\includegraphics[width=1.4in]{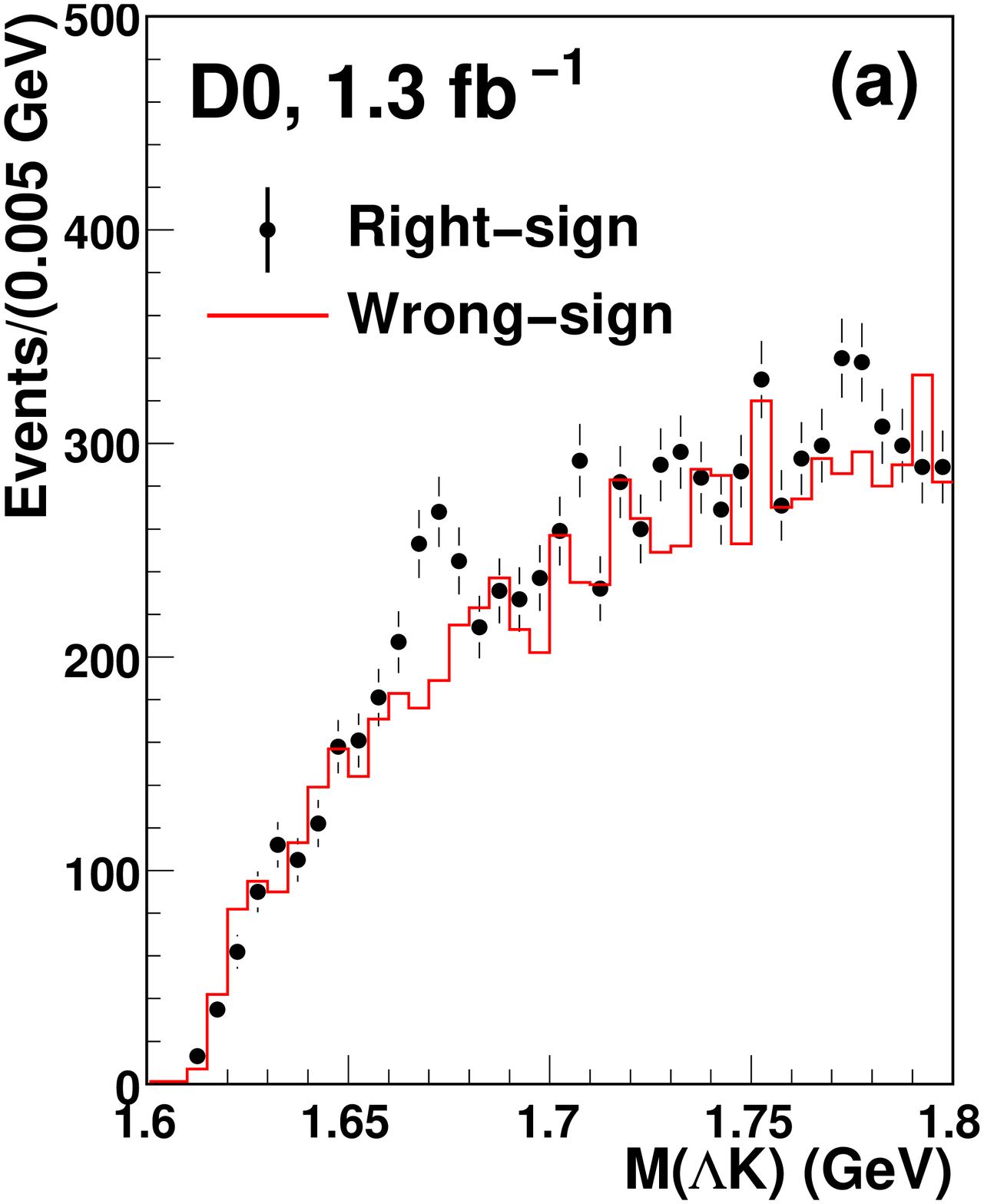} &  \includegraphics[width=1.4in]{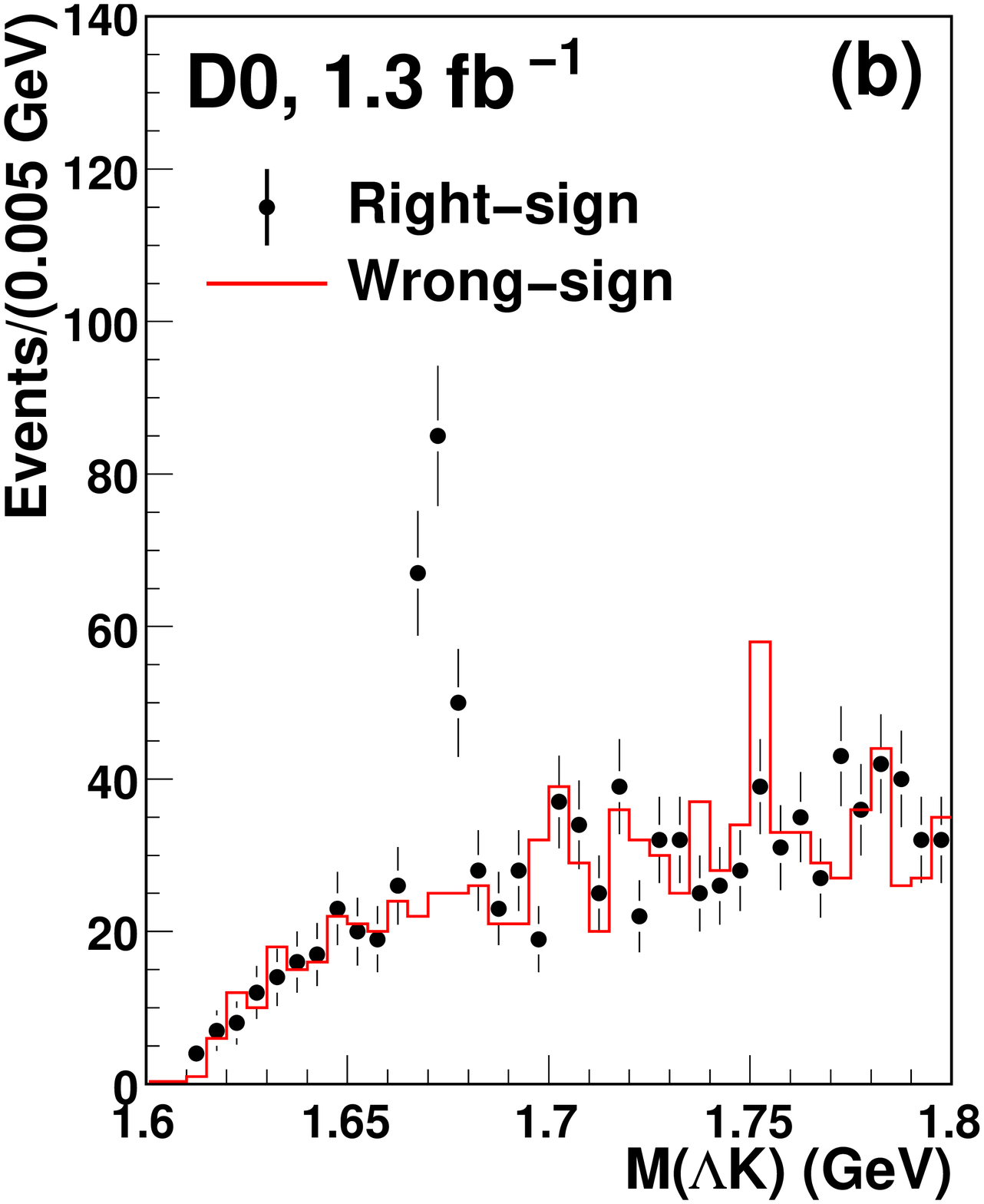} \\
\end{tabular}
\end{center}
\caption{The invariant mass distribution of the $\Lambda K$ pair before (a) and after (b) the BDT selection. Filled circles are from the right-sign $\Lambda K^-$ events while the histogram is from the wrong-sign $\Lambda K^+$ events without any additional normalization.}
\label{fig:Omega}
\end{figure}

We combine $\Lambda$ candidates with negatively charged particles (assumed to be kaons) to form $\Omega^-\to\Lambda K^-$ decay candidates. The $\Lambda$ and the kaon are required to have a common vertex. The $\Omega^-$ candidates must have transverse decay length significances greater than four and the uncertainties of the proper decay lengths less than 0.5~cm. These two requirements reduce backgrounds from combinatorics and mismeasured tracks. The $\Xi^{-}$ baryon has a mass of $1.322$~GeV~\cite{pdg} and decays into $\Lambda\pi^-$. If the kaon mass is assigned to the pion, this decay could be a major background for $\Omega^{-}\to\Lambda K^-$. To eliminate this background, we remove candidates with $\Lambda\pi^-$ mass less than $1.34$~GeV. Figure~\ref{fig:Omega}(a) shows the mass distribution of the reconstructed $\Omega^-\to\Lambda K^-$ candidates after these selections. The distribution of wrong-sign $\Lambda K^+$ events is also shown. An excess of events above the background around the expected $\Omega^-$ mass of $1.672$~GeV~\cite{pdg} is visible in the distribution of the right-sign $\Lambda K^-$ events. 

To further enhance the $\Omega^-$ signal over the combinatorial background, kinematic variables associated with daughter particle momenta, vertices, and track qualities are combined using Boosted Decision Trees (BDT)~\cite{tmva,BDT}. The $\Lambda K^-$ mass distribution after the BDT selection is shown in Fig.~\ref{fig:Omega}(b). The BDT selection retains 87\% of the $\Omega^-$ signal while rejecting 89\% of the background. The enhanced $\Omega^-$ mass peak is evident in the distribution. A $\Lambda K^-$ pair is considered to be a $\Omega^-$ candidate if its mass is in the range of $1.662-1.682$~GeV.
   
To select $\Omega_b^-\to J/\psi\thinspace\Omega^-$ candidates, we develop selection criteria using the MC $\Omega_b^-$ events as the signal and the data wrong-sign events as the background. The background events are formed by combining $J/\psi$ candidates with $\Lambda K^+$ pairs with mass between 1.662 and 1.682~GeV. We form $\Omega_b^-\to J/\psi\thinspace\Omega^-$ decay candidates from $J/\psi$ and $\Omega^-$ pairs that are consistent with being from a common vertex. We require the uncertainty of the $\Omega_b^-$ proper decay length to be less than 0.03~cm and impose a minimum $p_T$ cut of 6~GeV on the $\Omega_b^-$ candidates. Finally, $J/\psi$ and $\Omega^-$ daughters from the $\Omega_b^-$ decays are expected to be boosted in the direction of the $\Omega_b^-$; therefore, we require the opening angle in the transverse plane between the $J/\psi$ and the $\Omega^-$ to be less than $\pi/2$. 

\begin{figure}[htb]
\begin{center}
\mbox{\includegraphics[width=1.5in]{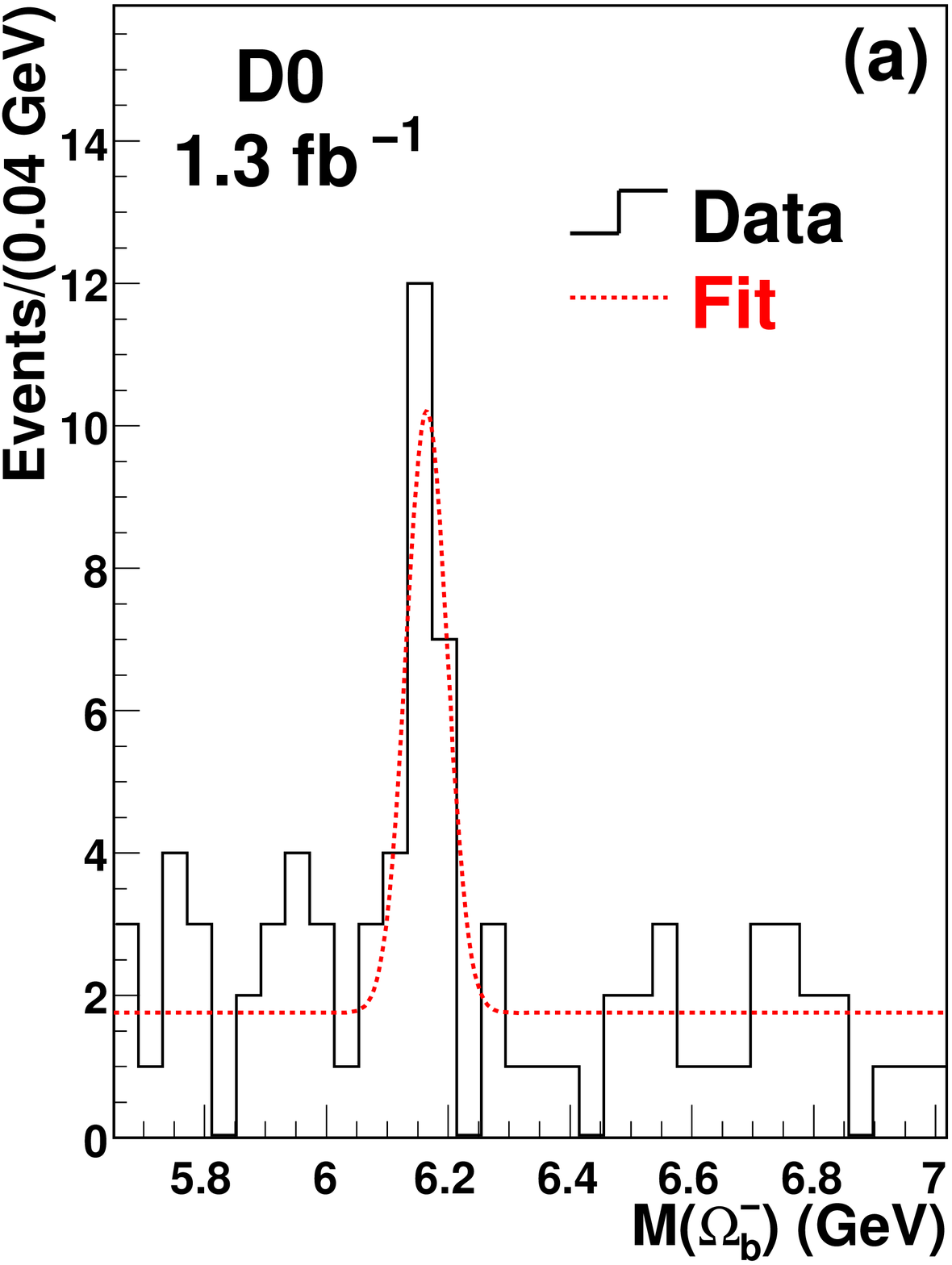}
      \includegraphics[width=1.57in]{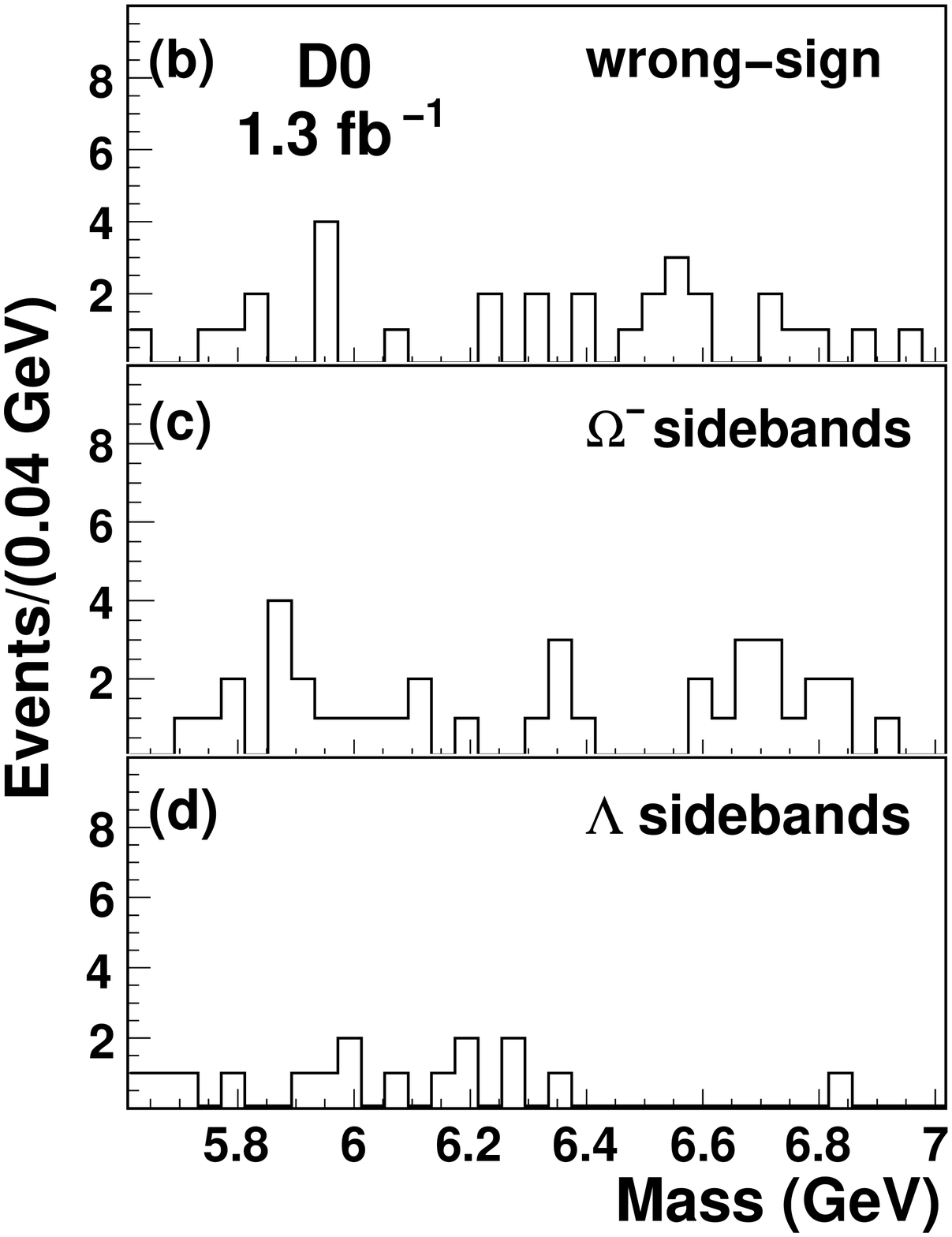}}
\end{center}
\caption{(a) The $M(\Omega_b^-)$ distribution of the $\Omega_b^-$ candidates after all selection criteria. The dotted curve is an unbinned likelihood fit to the model of a constant background plus a Gaussian signal. The mass distributions for the wrong-sign background~(b), the $\Omega^-$ sideband events~(c), and the $\Lambda$ sideband events~(d).} 
\label{fig:OmegabMass}
\end{figure}

We then apply the above selections to the right-sign events in the data to search for the $\Omega_b^-$ baryon in the mass window between $5.6$ and $7.0$~GeV. This range is chosen since 5.624~GeV is the mass of the lightest $b$ baryon, the $\Lambda_b$, and the upper limit of $7.0$~GeV is nearly 1 GeV higher than the predicted $\Omega_b^-$ mass~\cite{theory}. We calculate the $\Omega_b^-$ candidate mass using the formula $M(\Omega_b^-) = M(J/\psi\thinspace\Omega^-) - M(\mu^+\mu^-) - M(\Lambda K^-) + {\hat M}(J/\psi) + {\hat M}(\Omega^-)$. Here $M(J/\psi\thinspace\Omega^-)$, $M(\mu^+\mu^-)$, and $M(\Lambda K^-)$ are the reconstructed masses while ${\hat M}(J/\psi)$ and ${\hat M}(\Omega^-)$ are taken from Ref.~\cite{pdg}. This calculation improves the mass resolution of the MC $\Omega_b^-$ events from 0.080~GeV to $0.034$~GeV. In the mass search window, we observe 79 candidates in the data with the mass distribution shown in Fig.~\ref{fig:OmegabMass}(a). An excess of events near $6.2$~GeV is apparent. No such structure, however, is seen in the corresponding mass distribution (Fig.~\ref{fig:OmegabMass}(b)) of the 30 wrong-sign events. 

Assuming the excess is due to the $\Omega_b^-$ production, we fit $\Omega_b^-$ candidate masses with the hypothesis of a Gaussian signal plus a flat background using an unbinned likelihood method. We fix the Gaussian width to $0.034$~GeV, the width of the MC $\Omega_b^-$ signal. The fit gives an $\Omega_b^-$ mass of $6.165\pm 0.010\thinspace({\rm stat.})$~GeV and a yield of $17.8\pm 4.9\thinspace({\rm stat.})$ signal events. To assess the significance of the excess, we first determine the likelihood ${\cal L}_{s+b}$ of the signal plus background fit above and then repeat the fit with only the background contribution to find a new likelihood ${\cal L}_b$. The logarithmic likelihood ratio $\sqrt{2\ln({\cal L}_{s+b}/{\cal L}_b)}$ yields a statistical significance of 5.4$\sigma$, equivalent to a probability of $6.7\times 10^{-8}$ that the background could fluctuate with a significance equal to or greater than what is observed. Fitted yields for positively and negatively charged candidates are $6.2\pm 3.1\thinspace({\rm stat.})$ $\Omega_b^+$ and $12.0\pm 3.9\thinspace({\rm stat.})$ $\Omega_{b}^{-}$, respectively. 

Various checks have been performed to ensure that the observed resonance is genuine.  
(1) We apply the event selection to data events in the sidebands of the reconstructed $\Omega^-$ and $\Lambda$ resonances separately. The $J/\psi\thinspace(p\pi^-)\thinspace K^-$ mass distributions of these sideband events are shown in Figs.~\ref{fig:OmegabMass}(c) and~\ref{fig:OmegabMass}(d). No significant excess is present in either distribution.
(2) We investigate the possibility of a false signal due to residual $b$ hadron backgrounds by applying the final $\Omega_b^-$ selection to MC $B^-\to J/\psi\thinspace K^{*-}\to J/\psi\thinspace K_S^0\pi^-$, $\Xi^{-}_{b}\to J/\psi\thinspace \Xi^{-}$,  and $\Lambda_b\to J/\psi\thinspace\Lambda$ samples with equivalent integrated luminosities significantly greater than that of the analyzed data. No indication of any resonance is observed in the reconstructed $J/\psi\thinspace\Omega^-$ mass distribution. 
(3) We check the mass distributions of the $\Omega_{b}^{-}$ decay products.
For $\Omega_b^-$ candidates within a $\pm 3\sigma$ mass window around the observed peak, we relax the mass requirements on the $\Omega^{-}$ and $\Lambda$ candidates and perform a fit to each mass distribution. The numbers of the $\Omega^{-}$ and $\Lambda$ candidates from the fits are consistent with the observed number of $\Omega_{b}^{-}$ signal events.
(4) We replace the BDT selection with individual cuts on the most important variables according to the BDT optimization and confirm the existence of a peak with a comparable event yield but a higher background at a mass consistent with that observed using the BDT.
(5) We test the robustness of the peak by varying selections such as the $\Xi^{-}$ veto, $\Lambda$ and $\Omega^{-}$ mass windows, $\Lambda$ transverse decay requirements, BDT selection, and the requirement on $p_{T}(\Omega^{-}_{b})$. 
All the above studies confirm the existence of the resonance.

Potential sources of systematic uncertainties on the measured $\Omega_b^-$ mass include event selection, signal and background models, and momentum scale. Varying the selection criteria and applying a set of cuts on individual kinematic variables lead to a maximum change of 0.012~GeV in the fitted mass. Using a linear function as the background model results in negligible change in the mass. Varying the Gaussian width in the signal model between 0.028 and 0.040~GeV changes the fitted mass by at most 0.003~GeV. When a tighter selection is applied to enhance signal over background, we can float the width of the signal model in the fit. This leads to a mass shift of 0.002~GeV and a fitted signal width of 0.033 $\pm$ 0.010~GeV, consistent with the MC expectation. To study the effect of the track momentum scale uncertainty on the measured $\Omega_b^-$ mass, we reconstruct the higher statistics $\Lambda_{b}\to J/\psi\thinspace\Lambda$ decays and measure the $\Lambda_{b}$ mass for different minimum $p_T$ requirements on the $\Lambda_b$ daughter particles. We compare these measurements to the world average value of the $\Lambda_b$ mass~\cite{pdg} and take the maximum deviation of 0.004~GeV as a systematic uncertainty. Adding in quadrature, we get a total systematic uncertainty of 0.013~GeV to obtain a measured $\Omega_b^-$mass: $6.165\pm 0.010\thinspace({\rm stat.})\pm 0.013\thinspace({\rm syst.})$~GeV. Similarly, we estimate the systematic uncertainty on the $\Omega_b^-$ yield by varying the signal and background models in the fit. The observed maximum change of 0.8 is assigned as the systematic uncertainty on the yield: $17.8\pm 4.9\thinspace({\rm stat.})\pm 0.8\thinspace({\rm syst.})$. In all these studies, the signal significance remains above $5\sigma$.

\begin{figure}[htb]
\begin{center}
\includegraphics[width=2.5in]{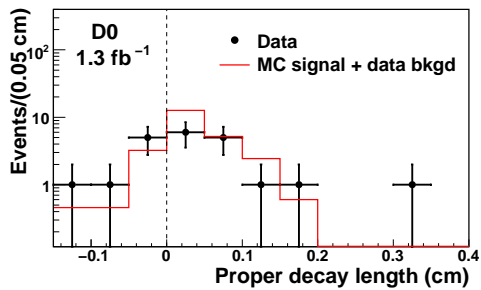}
\end{center}
\caption{The distribution of the proper decay length of the $\Omega_b^-$ candidates in the $\pm 3\sigma$ mass window around the observed peak along with the expected distribution from the MC $\Omega_b^-$ signal with a lifetime 1.54~ps plus the data background events.}
\label{fig:lifetime}
\end{figure}

Figure~\ref{fig:lifetime} shows the distribution of the proper decay length of the $\Omega_b^-$ candidates observed in the $\pm 3\sigma$ mass window around the fitted $\Omega_b^-$ mass. The distribution of the MC $\Omega_b^-$ signal plus the data background events is also shown. The background distribution is modeled using events in the $\Omega_b^-$ sidebands of $5.8-6.0$ and $6.4-6.6$~GeV. Despite the low statistics, the data distribution contains significantly more positive than negative decay lengths as expected and consistent with a weakly decaying $b$ hadron.

We calculate the $\Omega_b^-$ production rate relative to that of the $\Xi_b^-$~\cite{d0xib}. The selection efficiency ratio $\epsilon(\Omega_b^-\to J/\psi\thinspace\Omega^-)/\epsilon(\Xi_b^-\to J/\psi\thinspace\Xi^-)$ is found to be $1.5\pm 0.2\thinspace({\rm stat.})$ assuming inclusive $\Omega^-$ and $\Xi^-$ decays. The higher efficiency for the $\Omega_b^-$ is due primarily to a harder $p_T$ spectrum of the kaon from the $\Omega^-$ decay than that of the pion from the $\Xi^-$ decay and a shorter lifetime of the $\Omega^-$ compared to the $\Xi^-$. By using the reported $\Xi_b^{-}$ events~\cite{d0xib} and the observed $\Omega_{b}^{-}$ yield here, we estimate  
\begin{eqnarray*}
{\cal R}=\frac{f(b\to\Omega_b^-)}{f(b\to\Xi_b^-)}\cdot\frac{Br(\Omega_b^-\to J/\psi\thinspace\Omega^-)} {Br(\Xi_b^- \to J/\psi\thinspace \Xi^-)}
\end{eqnarray*}
to be ${\cal R} =0.80\pm 0.32\thinspace({\rm stat.})^{+0.14}_{-0.22}\thinspace({\rm syst.})$. Here $f(b\to\Omega_b^-)$ and $f(b\to\Xi_b^-)$ are the fractions of $b$ quarks that hadronize to form $\Omega_b^-$ and $\Xi_b^-$, respectively. The systematic uncertainty includes contributions from the signal yields as well as the efficiency ratio. Using $\Gamma(\Omega_b^-\to J/\psi\thinspace\Omega^-)/\Gamma(\Xi_b^-\to J/\psi\thinspace\Xi^-) = 9.8$~\cite{BRtheory}, the central values of $\tau(\Xi^{-}_{b})=1.42^{+0.28}_{-0.24}$~ps~\cite{pdg}, the ${\cal R}$ value above, and $\tau(\Omega_{b}^{-})$ in the range of $0.83-1.67$ ps~\cite{theory}, we obtain $f(b\to\Omega_b^-)/f(b\to\Xi_b^-)\approx 0.07-0.14$.

In summary, by analyzing 1.3 fb$^{-1}$ of data collected by the D0 experiment in $p\bar{p}$ collisions at $\sqrt{s}=1.96$ TeV at the Fermilab Tevatron Collider, we have made the first observation of the doubly strange $b$ baryon $\Omega_{b}^{-}$ in the fully reconstructed decay mode $\Omega_b^-\to J/\psi\thinspace\Omega^-$ with $J/\psi\to\mu^+\mu^-$, $\Omega^- \to\Lambda K^-$ and $\Lambda\to p\pi^-$. We measure the $\Omega_b^-$ mass to be $6.165\pm 0.010\thinspace({\rm stat.})\pm 0.013\thinspace({\rm syst.})$~GeV. The significance of the observed signal is greater than $5\sigma$.

% acknowledgement_paragraph_r2.tex                         7/25/08
%
We thank the staffs at Fermilab and collaborating institutions, 
and acknowledge support from the 
DOE and NSF (USA);
CEA and CNRS/IN2P3 (France);
FASI, Rosatom and RFBR (Russia);
CNPq, FAPERJ, FAPESP and FUNDUNESP (Brazil);
DAE and DST (India);
Colciencias (Colombia);
CONACyT (Mexico);
KRF and KOSEF (Korea);
CONICET and UBACyT (Argentina);
FOM (The Netherlands);
STFC (United Kingdom);
MSMT and GACR (Czech Republic);
CRC Program, CFI, NSERC and WestGrid Project (Canada);
BMBF and DFG (Germany);
SFI (Ireland);
The Swedish Research Council (Sweden);
CAS and CNSF (China);
and the
Alexander von Humboldt Foundation (Germany).

\end{document}